\documentclass[prl,showpacs,showkeys,nofootinbib,twocolumn,floatfix]{revtex4}

\usepackage{graphicx}  %
\usepackage{amsmath}
\usepackage{bm}  %
\usepackage{ulem}

\newcommand{\bea}{\begin{eqnarray}}
\newcommand{\eea}{\end{eqnarray}}
\newcommand{\beq}{\begin{equation}}
\newcommand{\eeq}{\end{equation}}
\newcommand{\bqa}{\begin{eqnarray}}
\newcommand{\eqa}{\end{eqnarray}}

\def\mqo2{{\!\!\!}}

\begin{document}

\title{Emission of Photons and Relativistic Axions\\
 from Axion Stars}

\author{Eric Braaten}
\email{braaten@mps.ohio-state.edu}
\affiliation{Department of Physics,
         The Ohio State University, Columbus, OH\ 43210, USA\\}

\author{Abhishek Mohapatra}
\email{mohapatra.16@buckeyemail.osu.edu}
\affiliation{Department of Physics,
         The Ohio State University, Columbus, OH\ 43210, USA\\}

\author{Hong Zhang}
\email{zhang.5676@osu.edu}
\affiliation{Department of Physics,
         The Ohio State University, Columbus, OH\ 43210, USA\\}

\date{\today}

\begin{abstract}
The number of nonrelativistic axions can be changed by inelastic reactions
that produce  photons or relativistic axions.
Any odd number of axions can annihilate into two photons.
Any even number of nonrelativistic axions can scatter into two relativistic axions.
We calculate the rate at which axions are lost from axion stars from these inelastic reactions.
In dilute systems of axions, the dominant inelastic reaction is axion decay into two photons.  
In sufficiently dense systems of axions, the dominant inelastic reaction
is the scattering of four nonrelativistic axions into two relativistic axions.
The scattering of odd numbers of axions into two photons produces
monochromatic radio-frequency signals at odd-integer harmonics of the fundamental frequency
set by the axion mass. 
This provides a unique signature for dense systems of axions, such as
a dense axion star or a collapsing dilute axion star.
\end{abstract}

\smallskip
\maketitle

{\bf Introduction}.
The particle nature of the  dark matter of the universe
remains one of the greatest mysteries in contemporary physics.
One of the most strongly motivated possibilities from a particle physics perspective
 is the {\it axion} \cite{Kim:2008hd},
which is the pseudo-Goldstone boson
associated with a $U(1)$ symmetry that solves the strong $CP$ problem of QCD.
The axion is a spin-0 particle with a very small mass
and extremely weak self-interactions.
Nonrelativistic axions with high occupation numbers can be produced in the early universe 
by a combination of the {\it cosmic string decay mechanism} \cite{Davis:1986xc,Harari:1987ht}
and the {\it vacuum misalignment mechanism} \cite{Preskill:1982cy,Abbott:1982af,Dine:1982ah}. 
The vacuum misalignment mechanism produces coherent axions.

A metastable gravitationally bound configuration of axions is called an {\it axion star}.
The ground state of an axion star is a Bose-Einstein condensate (BEC).
In the well-known solutions for axion stars,
the attractive forces from gravity and from the 4-axion interaction
are balanced by the kinetic pressure \cite{Barranco:2010ib,Chavanis:2011zm}.
We refer to these solutions as {\it dilute axion stars}, because the number density remains small enough 
that 6-axion and higher interactions are negligible.
There is a critical mass $M_*$ for the dilute axion star beyond which the kinetic pressure 
is unable to balance the attractive forces. 
There may also be {\it dense axion stars},
in which multi-axion interactions play an important role in the balance of forces \cite{Braaten:2015eeu}. 

Spacial fluctuations in the vacuum misalignment of the axion field in the early universe
naturally produce gravitationally bound ``miniclusters'' of axions with total mass in a range that includes
the critical mass $M_*$ of a dilute axion star \cite{Hogan:1988mp,Kolb:1993zz}.
Sikivie and collaborators have pointed out that gravitational interactions  
can  thermalize axion dark matter \cite{Sikivie:2009qn,Erken:2011dz}.
They drive an axion minicluster towards 
an axion star and then towards its BEC ground state.
Gravitational thermalization also allows the axion star to accrete more  axions.  
If accretion of axions increases the mass of the dilute axion star to above $M_*$,
it will collapse.  The fate of a collapsing axion star has not been established.
One possibility is that it collapses  to a black hole \cite{Chavanis:2016dab}.
Another possibility is a {\it bosenova}, in which  a burst of relativistic axions
is produced by inelastic axion reactions amplified by the increasing density \cite{Levkov:2016rkk}.
Such a phenomenon has been observed in collapsing BECs
of ultracold atoms \cite{JILA-bosenova}.
One possibility for the remnant after the collapse is a dense axion star \cite{Braaten:2015eeu,Eby:2016cnq}.

The number of nonrelativistic axions is almost conserved.
The number can be decreased by inelastic reactions in which
any odd number of axions annihilates into two photons
or any even number of axions scatters into two relativistic axions.
In this paper, we calculate the loss rate of axions from axion stars.
We also point out that monochromatic radio-frequency signals at odd-integer harmonics 
of a fundamental frequency provides a unique signature for dense systems of axions,
such as a collapsing dilute axion star or a dense axion star.

~

{\bf Relativistic axion field theory.}
Axions can be described by a relativistic quantum field theory with a real scalar field $\phi(x)$.
The Hamiltonian has the form
\begin{equation}
{\cal H} = \tfrac{1}{2}\dot{\phi}^2
+\tfrac{1}{2} \nabla \phi \cdot \nabla \phi
+ {\cal V}(\phi) .
\label{H-phi}
\end{equation}
The potential ${\cal V}$ is a periodic function of $\phi$ with period $2 \pi f_a$,
where $f_a$ is the axion decay constant.
The product $m_a f_a$ of the mass  of the axion  and its decay constant  is  
$(8 \times 10^7~{\rm eV})^2$ \cite{Kim:2008hd}.
Astrophysical and cosmological constraints restrict $f_a$
to the window between about $5 \times 10^{17}$~eV
and about $8 \times 10^{21}$~eV  \cite{Kim:2008hd}.
The window for $m_a$ is therefore
from about $10^{-6}$~eV to about $10^{-2}$~eV.
The expansion of ${\cal V}$ in powers of $\phi$ 
determines dimensionless  coupling constants $\lambda_{2j}$ for axion self-interactions:
\begin{equation}
{\cal V} = \frac12 m_a^2 \phi^2
+ m_a^2 f_a^2 \sum_{j=2}^{\infty} \frac{\lambda_{2j}}{(2j)!} \left( \frac{\phi}{f_a} \right)^{2j}.
\label{V-series}
\end{equation}
The relativistic axion potential ${\cal V}$  is produced by nonperturbative QCD effects, 
and it depends on quark mass ratios  \cite{diCortona:2015ldu}.
It can be approximated by the  {\it chiral potential}  \cite{DiVecchia:1980yfw}:
\begin{equation}
{\cal V} =
m_\pi^2 f_\pi^2 \left(1- \left[ 1- \frac{4 z}{(1+z)^2} \sin^2(\phi/2f_a) \right]^{1/2} \right),
\label{V-chiral}
\end{equation}
where $z = m_u/m_d$ is the up/down quark mass ratio.
For $z=0.48$, the first few coupling constants are $\lambda_4 = -0.34$,
$\lambda_6 = -0.13$, and $\lambda_8 = -0.87$.
A simple model for the relativistic potential is the {\it instanton potential}:
\begin{equation}
{\cal V}  = 
m_a^2 f_a^2\left[ 1 - \cos(\phi/f) \right],
\label{V-phi}
\end{equation}
for which the coupling constants are $\lambda_{2n} = (-1)^{n+1}$.

The Lagrangian for the coupling of the axion to the electromagnetic field is
\begin{equation}
{\cal L}_{\rm em} = 
\frac{c_{\rm em} \alpha }{16\pi f_a} 
\epsilon^{\mu\nu \alpha\beta}F_{\mu \nu} F_{\alpha \beta}  \phi.
\label{L-axionEM}
\end{equation}
where $\alpha \approx 1/137$ and 
$c_{\rm em}$ is a numerical coefficient that depends on the axion model \cite{Kim:2008hd}.
For example, $c_{\rm em} = -1.95$ for the simplest KSVZ model  \cite{Kim:1979if,Shifman:1979if}
and $c_{\rm em}=0.72$ in a simple DFSZ model \cite{Dine:1981rt,Zhitnitsky:1980tq}.
The decay rate of the axion into two photons is
\begin{equation}
\Gamma_a = 
\frac{c_{\rm em} ^2 \alpha ^2 m_a^3}{256 \pi^3 f_a^2}.
\label{Gamma}
\end{equation}
In the simplest KSVZ model with $m_a =10^{-4\pm2}$~eV,
the axion decay rate is $\Gamma_a = 6 \!\times\! 10^{-60 \pm 10}$~eV. 
(Here and below, upper and lower error bars in an exponent correspond to increasing 
and decreasing $m_a$ by two orders of magnitude from $10^{-4}$~eV.)
The axion lifetime is $3 \!\times\! 10^{36 \mp 10}$~years. 
This is tens of orders of magnitude larger than the lifetime of the universe, 
which is about $10^{10}$~years.

The relativistic axion potential implies that there are inelastic reactions 
that change the axion number, such as $(2j)a \to 2a$ with $j \ge 2$.
By the optical theorem, the rate for $(2j)a \to 2a$ 
is proportional to the imaginary part of the amplitude for $(2j)a \to (2j)a$ from two-axion cuts.
 Since each axion loop is suppressed by a factor of $m_a^2/f_a^2$,
which is $3 \!\times\! 10^{-48\pm 8}$, 
the rate for the reaction $(2j)a \to 2a$ is suppressed by $m_a^2/f_a^2$ 
compared to the rate for the elastic scattering process $(2j)a \to (2j)a$.
Inelastic reactions with more than two outgoing axions are suppressed
by more powers of $m_a^2/f_a^2$ and can be ignored.

~

{\bf Nonrelativistic axion effective field theory.}
Axions whose kinetic energies are much smaller than $m_a$ can be described by a 
nonrelativistic effective field theory called {\it axion EFT} 
with a complex scalar field $\psi(\bm{r},t)$ \cite{Braaten:2016kzc}. 
The effective Hamiltonian has the form
\begin{equation}
{\cal H}_{\rm eff} = 
\frac{1}{2m_a} \nabla \psi^* \cdot \nabla \psi
+ {\cal V}_{\rm eff}(\psi^* \psi) .
\label{Heff-psi}
\end{equation}
In the case of an axion BEC, 
the effective potential ${\cal V}_{\rm eff}$  gives 
the mean-field energy of the condensate as a function of its number density $\psi^* \psi$.
The expansion of the ${\cal V}_{\rm eff}$ in powers of $\psi^* \psi$
defines dimensionless  coupling constants $v_{j}$  for axion self-interactions:
\begin{equation}
{\cal V}_{\rm eff} = m_a \psi^* \psi 
+ m_a^2 f_a^2 \sum_{j=2}^{\infty} \frac{v_j}{(j!)^2} \left(\frac{\psi^* \psi}{2m_a f_a^2} \right)^j.
\label{Veff-series}
\end{equation}
The  coupling constants $v_j$ can be derived by matching low-energy 
scattering amplitudes at tree level in the relativistic theory and in axion EFT \cite{Braaten:2016kzc}.
The first few  coupling constants are  $v_2 = \lambda_4$, 
$v_3=\lambda_6 - 17 \lambda_4^2/8$, 
and $v_4 =\lambda_8 - 11 \lambda_4 \lambda_6 + 49 \lambda_4^3/4$
\cite{Braaten:2016kzc}.

\begin{figure}[t]
\centerline{ \includegraphics*[width=8cm,clip=true]{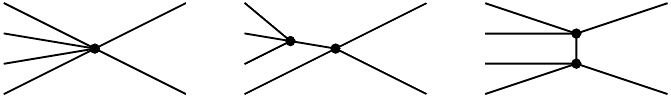} }
\vspace*{0.0cm}
\caption{Tree diagrams for $4a \to 2a$ in the relativistic axion theory. }
\label{fig:4to2tree}
\end{figure}

\begin{figure}[t]
\centerline{ \includegraphics*[width=8cm,clip=true]{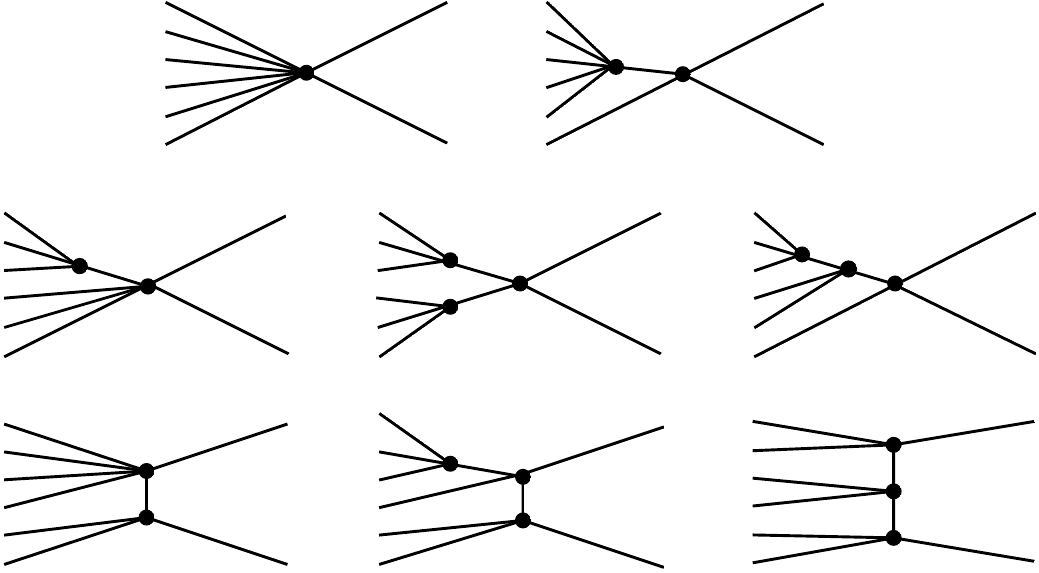} }
\vspace*{0.0cm}
\caption{Tree diagrams for $6a \to 2a$ in the relativistic axion theory. }
\label{fig:6to2tree}
\end{figure}

The Hamiltonian for axion EFT has a $U(1)$ phase symmetry. 
The associated conserved quantity is the number $N$ of low-energy axions defined by
\begin{equation}
N = \int \!\!d^3 r\, \psi^*\psi .
\label{Naxion-psi}
\end{equation}
Reactions in the relativistic axion theory that decrease the number of low-energy axions 
cannot be described explicitly within axion EFT, but their effects on nonrelativistic axions 
can be reproduced by the imaginary part of the effective potential,
which we denote by $-i {\cal W}_{\rm eff}$.
By the optical theorem, the rate for $(2j)a \to 2a$ with $j \ge 2$ is proportional to 
the imaginary part of the elastic scattering amplitude for  $(2j)a \to (2j)a$. 
The low-energy limit of the imaginary part of this amplitude is reproduced in axion EFT 
by the vertex from the $(\psi^* \psi)^{2j}$ term in ${\cal W}_{\rm eff}$.
Since ${\cal W}_{\rm eff}$ comes from matching one-loop diagrams in the relativistic theory,
it is suppressed by a factor of $m_a^2/f_a^2$ relative to ${\cal V}_{\rm eff}$.
It can be expanded in powers of $\psi^* \psi$:
\begin{equation}
{\cal W}_{\rm eff}  = 
m_a^4\sum_{j=2}^\infty \frac{w_{j+1}}{[(2j)!]^2}  
\left( \frac{\psi^* \psi}{2m_a f_a^2} \right)^{2j}.
\label{Weff-psi}
\end{equation}
The tree diagrams for $4a \to 2a$ and  $6a \to 2a$ in the relativistic theory 
are shown in Figs.~\ref{fig:4to2tree} and \ref{fig:6to2tree}, respectively. 
They determine the dimensionless coupling constants for the $(\psi^* \psi)^4$ and $(\psi^* \psi)^6$ terms 
 in Eq.~\eqref{Weff-psi}:
\begin{subequations}
\begin{eqnarray}
w_3 &=& \sqrt{3}\,[\lambda_6 - \lambda_4^2]^2/(64 \pi),
\\
w_4 &=& \sqrt{2}\,[\lambda_8  - \lambda_4 \lambda_6]^2/(48 \pi).
\end{eqnarray}
\end{subequations}
For the chiral potential in Eq.~\eqref{V-chiral} with $z=0.48$, 
these coefficients are $w_3 = 5.1 \!\times\! 10^{-4}$ and $w_4 = 7.9 \!\times\! 10^{-3}$. 
For the instanton potential in Eq.~\eqref{V-phi}, they are $w_3 = 0$ and $w_4 = 0$. 
We have verified that $w_5$ is also zero for the instanton potential.
We have no deep explanation for the vanishing of these coefficients.

\begin{figure}[t]
\centerline{ \includegraphics*[width=2.5cm,clip=true]{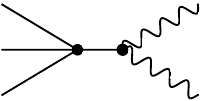} }
\vspace*{0.2cm}
\centerline{ \includegraphics*[width=6cm,clip=true]{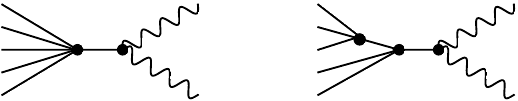} }
\vspace*{0.0cm}
\caption{Tree diagrams for $3 a \to \gamma \gamma$ and $5 a \to \gamma \gamma$ in the relativistic axion theory.} 
\label{fig:35to2gamma}
\end{figure}

The decay of the axion into two photons cannot be described explicitly within axion EFT,
because the energies of the photons in the axion rest frame are $m_a/2$,
which is beyond the range of validity of the low-energy effective theory.
However the effects of the decay on nonrelativistic axions 
can be reproduced by a term $-i \Gamma_a \psi^*\psi/2$ in the effective Hamiltonian density,
where $\Gamma_a$ is the decay rate in Eq.~\eqref{Gamma}.
In the relativistic theory,
any larger odd number $2j+1$ of axions can also annihilate into two photons. 
By the optical theorem, the rate for $(2j+1)a \to \gamma \gamma$ is proportional to 
the imaginary part of the amplitude for  $(2j+1)a \to (2j+1)a$ through a photon loop.
The low-energy limit of the imaginary part of this amplitude is  reproduced in axion EFT by 
the vertex from the $(\psi^* \psi)^{2j+1}$ term in an additional imaginary part of the effective potential
that we denote by $-i {\cal W}_{\rm em}$.
That potential can be expanded in powers of $\psi^* \psi$: 
\begin{equation}
{\cal W}_{\rm em}  = \tfrac12 \Gamma_a \psi^* \psi 
\left[ 1+  \sum_{j=1}^\infty \frac{u_{j+1}}{[(2j+1)!]^2}  
\left( \frac{\psi^* \psi}{2m_a f_a^2} \right)^{2j} \right].
\label{Wem-psi}
\end{equation}
The tree diagrams for $3 a \to \gamma \gamma$ and $5 a \to \gamma \gamma$
in the relativistic theory are shown in Fig.~\ref{fig:35to2gamma}.
They determine the dimensionless coupling constants for the $(\psi^* \psi)^3$ and $(\psi^* \psi)^5$ terms 
in Eq.~\eqref{Wem-psi}:
\begin{subequations}
\begin{eqnarray}
u_2 &=& 81 \lambda_4^2/64,
\\
u_3 &=&625 [ \lambda_6 + 5 \lambda_4^2/4]^2/576.
\end{eqnarray}
\end{subequations}
For the chiral potential with $z=0.48$, 
these coefficients are $u_2 = 0.15$ and $u_3 = 4.8 \!\times\! 10^{-4}$. 
For the instanton potential, they are $u_2 = 1.3$ and $u_3 = 5.5$.

The imaginary parts of the effective potential,  ${\cal W}_{\rm eff}$
in Eq.~\eqref{Weff-psi} and  and ${\cal W}_{\rm em}$ in Eq.~\eqref{Wem-psi},  
are functions of $\psi^* \psi$, so they are invariant under the $U(1)$ symmetry.
The effective Hamiltonian obtained by adding $-i({\cal W}_{\rm eff} + {\cal W}_{\rm em})$
to the effective potential ${\cal V}$ in Eq.~\eqref{Heff-psi} therefore
commutes with the number operator $N$ in Eq.~\eqref{Naxion-psi}.
This may seem to suggest that the time evolution generated by 
the effective Hamiltonian  conserves the number $N$ of low-energy axions.
However it is intuitively obvious that
the number of low-energy axions must decrease  with  time
as inelastic reactions convert nonrelativistic axions into pairs of relativistic axions
or into pairs of photons.  The resolution to this puzzle is that the effective density matrix 
for low-energy axions evolves in time according to a Lindblad equation,
with local Lindblad operators that are determined by the local
anti-Hermitian terms in the effective Hamiltonian  \cite{Braaten:2016sja}.

Axions can be lost from an axion star by the inelastic 
reactions $(2j)a \to 2a$ for any $j=2,3,\ldots$
and $(2j+1)a \to \gamma \gamma$ for any $j=0,1,2,\ldots$.
The local rate of decrease in the number density $n= \psi^*\psi$ 
of a BEC of nonrelativistic axions from $(2j)a \to 2a$ is given by
the $(\psi^* \psi)^{2j}$ term in ${\cal W}_{\rm eff}$ in Eq.~\eqref{Weff-psi} multiplied by $2j$. 
The local rate of decrease in $n$ from  $(2j+1)a \to \gamma \gamma$ is given by
the $(\psi^* \psi)^{2j+1}$ term in ${\cal W}_{\rm em}$ in Eq.~\eqref{Wem-psi} multiplied by $2j+1$. 
The total loss rate for nonrelativistic axions is obtained by adding all the terms 
in the local loss rate and integrating over the volume of the star:
\begin{eqnarray}
\frac{1}{N}\frac{dN}{dt} &=& 
- \frac{m_a^3}{f_a^2} \sum_{j=2}^\infty 
\frac{2j w_{j+1} \, \langle n^{2j-1} \rangle}{[(2j)!]^2 (2m_a f_a^2)^{2j-1}}
\nonumber\\
&&- \Gamma_a \bigg( 1
+ \sum_{j=1}^\infty \frac{(2j+1)u_{j+1} \, \langle n^{2j} \rangle}{[(2j+1)!]^2  (2m_a f_a^2)^{2j}}  
\Bigg),
\label{dN/dt}
\end{eqnarray}
where $ \langle n^k \rangle$ is the density-weighted average of a power of the 
number density:
$ \langle n^k \rangle= \int \!\! d^3r \, n^{k+1}/N$.

In Ref.~\cite{Eby:2015hyx}, the authors proposed a mechanism for the decay of axion stars
in which three axions from a BEC make a transition to a single relativistic axion. 
The reaction $3a \to a$ violates  conservation of energy and momentum.
The authors suggested that momentum conservation can be ignored,
because a tiny recoil momentum of the entire star would 
allow momentum to be conserved.
However there is no physical mechanism that can effectively transfer momentum 
from the few axions participating in the reaction to the many axions that make up the star.
That the $3a \to a$ reaction is not a viable mechanism for the loss of low-energy axions
is clear from axion EFT.  In the low-energy limit, the amplitude for $3a \to a \to 3a$ in the relativistic theory 
is reproduced in axion EFT by the coupling constant $v_3$ 
for the $(\psi^* \psi)^3$ term in ${\cal W}_{\rm eff}$ in Eq.~\eqref{Weff-psi}.  
The coupling constant for this operator has an imaginary part proportional to $u_2$
from the  $3 a \to \gamma \gamma$ reaction, but it
has no imaginary part that would correspond to a $3 a \to a$ reaction.
That the reaction $3a \to a$ proposed in Ref.~\cite{Eby:2015hyx} 
is not a viable mechanism for the emission of axions from a BEC
is also clear from experiments on BEC's of ultracold atoms \cite{Wieman97,Ketterle98,Grimm03}.
The dominant mechanism for the loss of atoms is usually the 3-body recombination reaction
$3a \to (aa) + a$, where $(aa)$ represents a diatomic molecule  \cite{Braaten:2016dsw}.
This reaction conserves energy and momentum.

~ 

{\bf Dilute axion stars.}
 Approximate solutions for {\it dilute axion stars} were first found numerically
by Barranco and Bernal \cite{Barranco:2010ib}
using the relativistic axion field theory with the instanton potential ${\cal V}$  in Eq.~\eqref{V-phi} 
and with gravitational interactions described by general relativity.
Accurate solutions were obtained more simply by Chavanis and Delfini
using a nonrelativistic axion field theory with the effective potential
${\cal V}_{\rm eff}$ truncated after the $(\psi^*\psi)^2$ term
and with gravitational interactions described by Newtonian gravity \cite{Chavanis:2011zm}.
They obtained an analytic result for the critical mass $M_*$
above which there are no stable spherically symmetric solutions: 
\begin{equation}
M_* = 10.15~ |\lambda_4|^{-1/2}f_a/\sqrt{G m_a^2}.
\label{Mmax-axion}
\end{equation}
If  $m_a = 10^{-4\pm 2}$~eV, the critical mass is   for the chiral potential with $z = 0.48$
is $10^{-13\mp 4}\,M_\odot$, where $M_\odot$ is the solar mass.
The solutions for axion stars are conveniently parametrized by the central number density $n_0$. 
The critical value of the  dimensionless central density
$\hat n_0 = n_0/(\tfrac12 m_af_a^2)$ is $800 \lambda_4^{-2} G f_a^2$ \cite{Chavanis:2011zm}, 
which is  $2\times 10^{-13 \mp4}$   for the chiral potential with $z = 0.48$.
We refer to the branch of axions stars with masses extending up to the critical mass
$M_*$ in Eq.~\eqref{Mmax-axion} as {\it dilute axion stars},
because the number density is always small enough that ${\cal V}_{\rm eff}$
in Eq.~\eqref{Veff-series} can be truncated after the $(\psi^* \psi)^2$ term. 
In a dilute axion star, the kinetic pressure of the axions balances the attractive forces
from gravity and from axion pair interactions.

\begin{figure}[t]
\centerline{ \includegraphics*[width=8.5cm,clip=true]{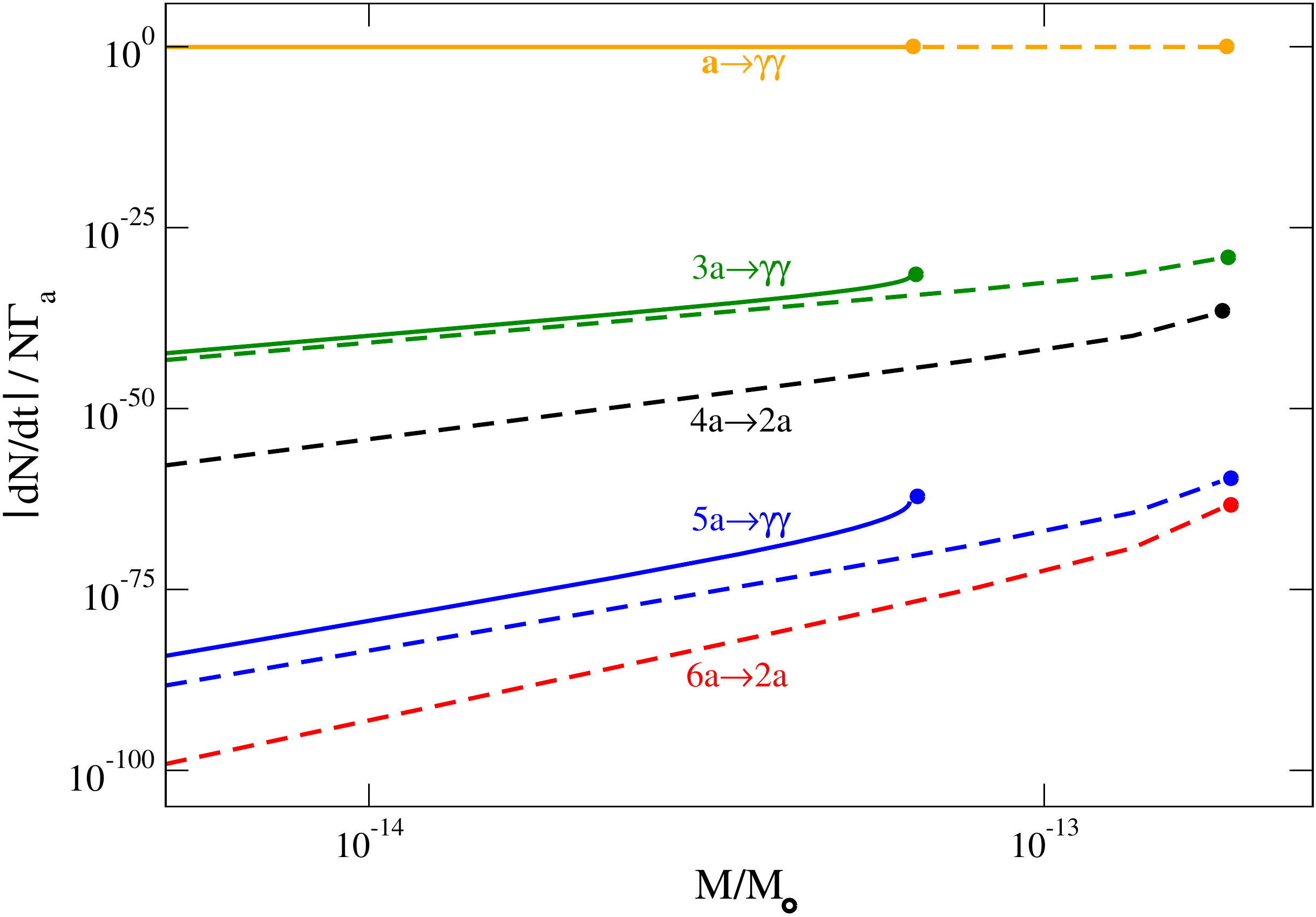} }
\vspace*{0.0cm}
\caption{
Axion loss rates from a  dilute axion star as functions of the mass $M$ of the axion star
(normalized to $M_\odot$).
The fractional loss rates $|dN/dt|/N$ for $m_a = 10^{-4}$~eV 
are normalized to the axion decay rate $\Gamma_a = 6.2 \times10^{-60}$~eV.
The loss rates are for the chiral potential with $z=0.48$ (dashed lines)
and for the instanton potential (solid lines).
}
\label{fig:ratio-dilute}
\end{figure}

\begin{figure}[t]
\centerline{ \includegraphics*[width=8.5cm,clip=true]{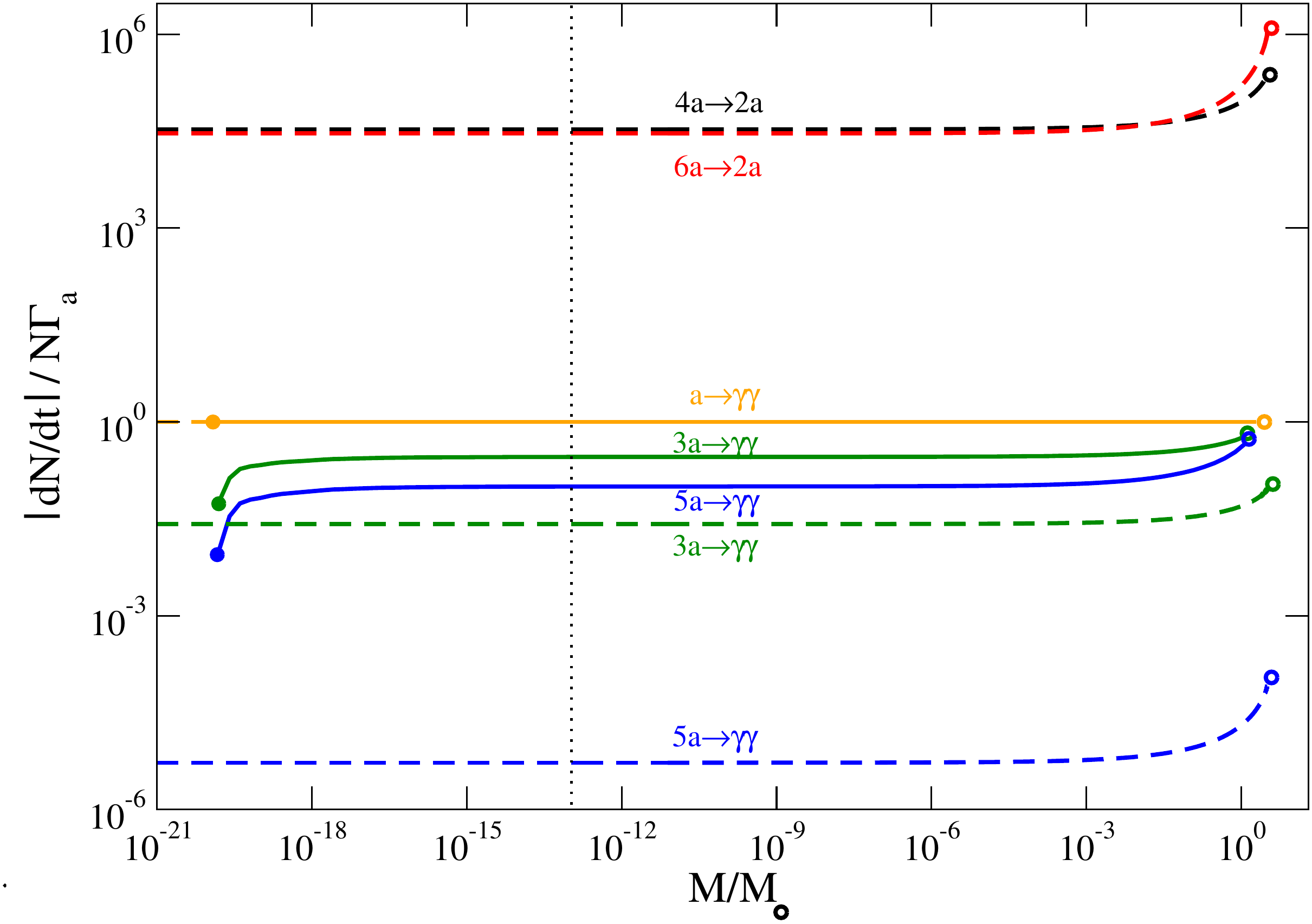} }
\vspace*{0.0cm}
\caption{
Axion loss rates from a dense axion star as functions of the mass $M$ of the axion star
(normalized to $M_\odot$).
The fractional loss rates $|dN/dt|/N$ for $m_a = 10^{-4}$~eV 
are normalized to the axion decay rate $\Gamma_a = 6.2 \times10^{-60}$~eV.
The loss rates are for the chiral potential with $z=0.48$ 
in the Thomas-Fermi approximation (dashed lines)
and for the instanton potential (solid lines).
The vertical dotted line is the critical mass for a dilute axion star with the instanton potential.
}
\label{fig:ratio-dense}
\end{figure}

The dominant contribution to the loss rate of axions from a dilute axion star
is from the decay $a \to \gamma \gamma$.
The lifetime of the dilute axion star is therefore the same as the lifetime $1/\Gamma_a$ of the axion.
Other individual contributions to the axion loss rate in Eq.~\eqref{dN/dt} 
are shown in Fig.~\ref{fig:ratio-dense} as functions of the mass $M$ of the dilute axion star.
At the critical mass $M_*$, the loss rate from $3a \to \gamma \gamma$
is suppressed by about $10^{-32}$. 
For the chiral potential, 
the loss rate at $M_*$ from $4a \to 2a$ is suppressed by about $10^{-47}$.
For the instanton potential, the operators in Eq.~\eqref{Weff-psi}  
give no contribution to the  loss rate from $(2j)a \to 2a$, at least for $j=2,3,4$.
For $j=2$, the largest contribution  
comes from gradient operators, such as $\psi^*\psi \nabla\psi^* \!\cdot\! \nabla \psi$.
The two gradients give an additional suppression factor of $|\mu|/m_a$, 
where $\mu$ is the chemical potential.
The chemical potential at the critical mass $M_*$ is
$-36 |\lambda_4|^{-1} G m_a f_a^2$ \cite{Chavanis:2011zm}.  
Thus the suppression factor $|\mu|/m_a$ is  about $10^{-17 \mp4}$  at $M_*$. 
As the mass $M$ decreases, the fractional loss rates in Fig.~\ref{fig:ratio-dense} decrease.
For $M \ll M_*$, the fractional loss rate from  $3a \to \gamma \gamma$ decreases 
roughly as $M^{8.3}$ for the chiral potential and $M^{8.7}$ for the instanton potential.
The fractional loss rate from  $4 a \to 2a$ decreases roughly as $M^{12.4}$ for the chiral potential. 

~

{\bf Dense axion stars.}
The possibility of  other branches of much denser axion stars 
has recently been suggested \cite{Braaten:2015eeu}.
In a {\it dense axion star},  the number density $\psi^*\psi$ becomes too large for the
effective potential ${\cal V}_{\rm eff}$  to
be approximated by a truncation of its expansion in powers of $\psi^* \psi$.
The attractive force of gravity and the repulsive force from the kinetic pressure
are balanced by the mean-field pressure of the axion BEC.
In Ref.~\cite{Braaten:2016kzc}, a systematically improvable sequence of approximations 
to the effective potential of  axion EFT that resum terms with all powers in of $\psi^* \psi$
was developed.  For the instanton potential in Eq.~\eqref{V-phi},
the first in this sequence of approximations to ${\cal V}_{\rm eff}$ is
\begin{equation}
{\cal V}_{\rm eff}^{(0)} =  \tfrac12  m_a \psi^* \psi
+m_a^2 f_a^2  \big[ 1 - J_0( \hat n^{1/2}) \big] ,
\label{Vclass-Bessel}
\end{equation}
where $\hat n = \psi^* \psi/(\tfrac12 m_af_a^2)$. This effective potential was derived previously
by using a nonrelativistic reduction of the relativistic axion field theory \cite{Eby:2014fya}.
For the chiral potential in Eq.~\eqref{V-chiral},
there is no analytic expression for the effective potential analogous to Eq.~\eqref{Vclass-Bessel}.
In Ref.~\cite{Braaten:2015eeu}, the differential equations for axion EFT 
with the instanton effective potential in Eq.~\eqref{Vclass-Bessel} and with gravitational interactions 
described by Newtonian gravity were solved to obtain solutions for dense axion stars.
For $m= 10^{-4\pm2}$~eV, the branch of dense axion stars begins at a lower
critical mass  $1.2 \times 10^{-20\mp 6}M_\odot$.
The lower critical mass for the chiral potential is not known. 
For the instanton potential, the dimensionless central density $\hat n_0$  is 13 at the lower critical mass.
The accurate numerical solution of  the differential equations for the axion star
becomes increasingly challenging as $\hat n_0$  increases. 
The mean-field energy of the axion BEC also becomes increasingly large compared 
to the kinetic energy of the axions.
The solution can therefore be simplified using the {\it Thomas-Fermi approximation}, 
in which the kinetic energy term for the axion field in
the differential equations for the axion star is neglected \cite{Wang:2001wq}.
The Thomas-Fermi approximation was used in Ref.~\cite{Braaten:2015eeu}
to extend the branch of dense axion stars up to the upper endpoint $1.9\, M_\odot$, 
beyond which there is no solution within this approximation.
For the effective chiral potential analogous to Eq.~\eqref{Vclass-Bessel},
it is relatively easy to calculate the solutions for dense axion stars
using the Thomas-Fermi approximation.

The individual contributions to the axion loss rate in Eq.~\eqref{dN/dt} from a dense axion star
are shown in Fig.~\ref{fig:ratio-dense} as functions of the mass $M$ of the axion star.
For the chiral  effective potential analogous to Eq.~\eqref{Vclass-Bessel} with $z=0.48$,
the solutions for dense axion stars were calculated using the Thomas-Fermi approximation.
This approximation begins to break down a few orders of magnitude above the lower critical mass. 
Since the Thomas-Fermi approximation does not predict such a critical point,
the fractional loss rates in Fig.~\ref{fig:ratio-dense} for the chiral potential have been extended 
all the way to the left end of the plot.
The fractional loss rates in Fig.~\ref{fig:ratio-dense} 
increase very slowly over  most of the range of $M$.
That range includes the critical mass $M_*$ for a dilute axion star.
For the chiral potential with $z=0.48$, the largest loss rate is from the reaction $4a \to 2a$.
At the critical mass $M_* $ for a dilute axion star,
the  loss rate from $6a \to 2a$ is smaller by a factor of $0.9$
and the loss rate from $a \to \gamma \gamma$ is smaller by a factor of $3 \times 10^{-5}$. 
For the instanton potential, 
the largest loss rate from producing relativistic axions
may come from a gradient operator, such as $\psi^*\psi \nabla\psi^* \!\cdot\! \nabla \psi$.
At $M_*$, this loss rate has Thomas-Fermi suppression
on top of the suppression by $|\mu|/m_a$, which is about $10^{-16 \mp 4}$.
Even if the loss rates from the  reactions $(2j+1)a \to \gamma \gamma$
are orders of magnitude smaller than the loss rates from $(2j)a \to 2a$, 
these reactions may still be important because
they produce monochromatic photons in the radio-frequency range that could be observed.
At the critical mass $M_*$ for a dilute axion star,
the fractional loss rates for a dense axion star from $3a \to \gamma \gamma$ and from $5a \to \gamma \gamma$ 
are smaller than  $\Gamma_a$ by $3 \times 10^{-2}$ and by $5 \times 10^{-6}$ for the chiral potential.
The fractional loss rates from $3a \to \gamma \gamma$  and from $5a \to \gamma \gamma$ 
are smaller than  $\Gamma_a$ by $0.29$  and by $0.10$  for the instanton potential. 

~

{\bf Odd-integer harmonics.}
One of the most puzzling discoveries in astrophysics in recent decades
is {\it fast radio bursts} \cite{Katz:2016dti}.
There are proposed mechanisms for fast radio bursts involving
monochromatic radio-frequency signals from dilute axion stars.
Iwazaki suggested that a signal with frequency $m_a$
could be produced  in the collision of a dilute axion star and a neutron star by
coherent electric-dipole radiation from electrons in the atmosphere of the neutron star \cite{Iwazaki:1999fd}.
Tkachev suggested that a signal with frequency $m_a/2$
could be produced by  a maser mechanism in a collapsing dilute axion star \cite{Tkachev:2014dpa}.
Raby recently suggested that a signal with frequency $m_a$
could be produced in the collision of a dilute axion star and a neutron star by
coherent electric-dipole radiation from neutrons in the outer core of the neutron star \cite{Raby:2016deh}.
With any of these mechanisms, odd-integer harmonics of the fundamental frequency
should also be produced, but at rates smaller by tens of orders of magnitude.
For the analogous mechanisms involving a dense axion star or any other dense systems of axions, 
each successive odd-integer harmonic may only be smaller by one or two orders of magnitude.
While cosmological and gravitational red shifts can change the fundamental frequency 
determined by $m_a$,  they will not affect the odd-integer ratios of the harmonics.
Thus monochromatic radio-frequency signals at odd-integer harmonics 
of a fundamental frequency are a unique signature for dense axion stars and
collapsing dilute axion stars.

\newpage 

\begin{acknowledgments}
This research was supported in part by the
Department of Energy under the grant DE-SC0011726
and by the National Science Foundation under the grant PHY-1310862.
HZ thanks K.~Ng and S.~Li for many beneficial discussions.

\end{acknowledgments}


\begin{thebibliography}{99}


\bibitem{Kim:2008hd} 
  J.~E.~Kim and G.~Carosi,
Axions and the strong CP problem,
  Rev.\ Mod.\ Phys.\  {\bf 82}, 557 (2010)
  [arXiv:0807.3125].
  
\bibitem{Davis:1986xc} 
  R.L.~Davis,
Cosmic axions from cosmic strings,
  Phys.\ Lett.\ B {\bf 180}, 225 (1986).
  
\bibitem{Harari:1987ht} 
  D.~Harari and P.~Sikivie,
On the Evolution of global strings in the early universe,
  Phys.\ Lett.\ B {\bf 195}, 361 (1987).

\bibitem{Preskill:1982cy} 
  J.~Preskill, M.B.~Wise and F.~Wilczek,
Cosmology of the invisible axion,
  Phys.\ Lett.\ B {\bf 120}, 127 (1983).
  
\bibitem{Abbott:1982af} 
  L.F.~Abbott and P.~Sikivie,
A cosmological bound on the invisible axion,
  Phys.\ Lett.\ B {\bf 120}, 133 (1983).

\bibitem{Dine:1982ah} 
  M.~Dine and W.~Fischler,
The not so harmless axion,
  Phys.\ Lett.\ B {\bf 120}, 137 (1983).
  
\bibitem{Barranco:2010ib} 
  J.~Barranco and A.~Bernal,
Self-gravitating system made of axions,
  Phys.\ Rev.\ D {\bf 83}, 043525 (2011)
  [arXiv:1001.1769].
    
\bibitem{Chavanis:2011zm} 
P.H.~Chavanis and L.~Delfini,
Mass-radius relation of Newtonian self-gravitating Bose-Einstein condensates with short-range interactions: 
II. Numerical results,
  Phys.\ Rev.\ D {\bf 84}, 043532 (2011)
  [arXiv:1103.2054].
  
\bibitem{Braaten:2015eeu} 
  E.~Braaten, A.~Mohapatra and H.~Zhang,
Dense axion stars,
  arXiv:1512.00108.
  
\bibitem{Hogan:1988mp} 
  C.J.~Hogan and M.J.~Rees,
Axion miniclusters,
  Phys.\ Lett.\ B {\bf 205}, 228 (1988).
  
\bibitem{Kolb:1993zz} 
  E.W.~Kolb and I.I.~Tkachev,
Axion miniclusters and Bose stars,
  Phys.\ Rev.\ Lett.\  {\bf 71}, 3051 (1993)
  [hep-ph/9303313].
  
\bibitem{Sikivie:2009qn} 
  P.~Sikivie and Q.~Yang,
Bose-Einstein condensation of dark matter axions,
  Phys.\ Rev.\ Lett.\  {\bf 103}, 111301 (2009)
  [arXiv:0901.1106].
  
\bibitem{Erken:2011dz} 
  O.~Erken, P.~Sikivie, H.~Tam, and Q.~Yang,
Cosmic axion thermalization,
  Phys.\ Rev.\ D {\bf 85}, 063520 (2012)
  [arXiv:1111.1157].
  
\bibitem{Chavanis:2016dab} 
  P.H.~Chavanis,
Collapse of a self-gravitating Bose-Einstein condensate with attractive self-interaction,
  arXiv:1604.05904.
  
\bibitem{JILA-bosenova} 
E.A. Donley, N.R.~Claussen, S.L.~Cornish, J.L.~Roberts, E.A.~Cornell, and C.E.~Wieman,
Dynamics of collapsing and exploding Bose-Einstein condensates,
Nature {\bf 412}, 295 (2001)
[cond-mat/0105019].

\bibitem{Eby:2016cnq} 
  J.~Eby, M.~Leembruggen, P.~Suranyi, and L.C.R.~Wijewardhana,
Collapse of axion stars,
  arXiv:1608.06911.

\bibitem{Levkov:2016rkk} 
  D.G.~Levkov, A.G.~Panin and I.I.~Tkachev,
Relativistic axions from collapsing Bose stars,
  arXiv:1609.03611.
  
\bibitem{DiVecchia:1980yfw} 
  P.~Di Vecchia and G.~Veneziano,
Chiral dynamics in the large $N$ limit,
  Nucl.\ Phys.\ B {\bf 171}, 253 (1980).

\bibitem{diCortona:2015ldu} 
  G.G.~di Cortona, E.~Hardy, J.P.~Vega and G.~Villadoro,
The QCD axion, precisely,
   JHEP {\bf 1601}, 034 (2016)
  [arXiv:1511.02867].
  
\bibitem{Kim:1979if} 
  J.E.~Kim,
Weak interaction singlet and strong $CP$ invariance,
  Phys.\ Rev.\ Lett.\  {\bf 43}, 103 (1979).

\bibitem{Shifman:1979if} 
  M.A.~Shifman, A.I.~Vainshtein and V.I.~Zakharov,
Can confinement ensure natural $CP$ invariance of strong interactions?,
  Nucl.\ Phys.\ B {\bf 166}, 493 (1980).

\bibitem{Dine:1981rt} 
  M.~Dine, W.~Fischler and M.~Srednicki,
A simple solution to the strong $CP$ problem with a harmless axion,
  Phys.\ Lett.\ B {\bf 104}, 199 (1981).

\bibitem{Zhitnitsky:1980tq} 
 A.R.~Zhitnitsky,
On possible suppression of the axion hadron interactions,
  Sov.\ J.\ Nucl.\ Phys.\  {\bf 31}, 260 (1980)
  [Yad.\ Fiz.\  {\bf 31}, 497 (1980)].

\bibitem{Braaten:2016kzc} 
  E.~Braaten, A.~Mohapatra and H.~Zhang,
Nonrelativistic effective field  theory for axions,
  arXiv:1604.00669.

\bibitem{Braaten:2016sja} 
  E.~Braaten, H.-W.~Hammer and G.P.~Lepage,
Open effective field theories from deeply inelastic reactions,
  arXiv:1607.02939 [hep-ph].
      
\bibitem{Eby:2015hyx} 
  J.~Eby, P.~Suranyi and L.~C.~R.~Wijewardhana,
The lifetime of axion stars,
  Mod.\ Phys.\ Lett.\ A {\bf 31}, 1650090 (2016)
  [arXiv:1512.01709].
  
\bibitem{Wieman97}
E.A.~Burt, R.W.~Ghrist, C.J.~Myatt, M.J.~Holland, E.A.~Cornell, and C.E.~Wieman,
Coherence, correlations, and collisions: What one learns about Bose-Einstein condensates from their decay,
Phys.\ Rev.\ Lett.\ {\bf 79}, 337 (1997).

\bibitem{Ketterle98}
D.M.~Stamper-Kurn, M.R.~Andrews, A.P.~Chikkatur, S.~Inouye, H.-J.~Miesner, J.~Stenger, and W.~Ketterle,
Optical confinement of a Bose-Einstein condensate,
Phys.\ Rev.\ Lett.\ {\bf 80}, 2027 (1998).

\bibitem{Grimm03}
T.~Weber, J.~Herbig, M.~Mark, H-C.~N\"agerl, and R.~Grimm,
Three-body recombination at large scattering lengths in an ultracold atomic gas,
Phys.\ Rev.\ Lett.\ {\bf 91}, 123201 (2003).

\bibitem{Braaten:2016dsw} 
  E.~Braaten, H.-W.~Hammer and G.P.~Lepage,
Lindblad equation for the inelastic loss of ultracold atoms,
  arXiv:1607.08084.
  
\bibitem{Eby:2014fya} 
  J.~Eby, P.~Suranyi, C.~Vaz and L.C.R.~Wijewardhana,
Axion stars in the infrared limit,
   JHEP {\bf 1503}, 080 (2015)
  [arXiv:1412.3430].
   
\bibitem{Tkachev:1991ka} 
  I.I.~Tkachev,
On the possibility of Bose star formation,
  Phys.\ Lett.\ B {\bf 261}, 289 (1991).

\bibitem{Wang:2001wq} 
  X.Z.~Wang,
Cold Bose stars: self-gravitating Bose-Einstein condensates,
  Phys.\ Rev.\ D {\bf 64}, 124009 (2001).
  
\bibitem{Katz:2016dti} 
  J.~I.~Katz,
Fast radio bursts -- A brief review: Some questions, fewer answers,
  Mod.\ Phys.\ Lett.\ A {\bf 31}, 1630013 (2016)
  [arXiv:1604.01799].

\bibitem{Iwazaki:1999fd} 
  A.~Iwazaki,
Emission of radio waves in gamma-ray bursts and axionic boson stars,
  hep-ph/9908468.

\bibitem{Tkachev:2014dpa} 
  I.I.~Tkachev,
Fast radio bursts and axion miniclusters,
  JETP Lett.\  {\bf 101}, 1 (2015)
  [arXiv:1411.3900].

\bibitem{Raby:2016deh} 
  S.~Raby,
Axion star collisions with neutron stars and fast radio bursts,
   arXiv:1609.01694.
  
\end{thebibliography}
\end{document}